\documentclass[preprint]{revtex4}
\usepackage{bm}
\input{epsf}
\begin{document}
\title{Strain induced band alignment in wurtzite-zincblende InAs heterostructured nanowires}

\author{Jaya Kumar Panda and Anushree Roy}
\email{anushree@phy.iitkgp.ernet.in} \affiliation{Department of
Physics, Indian Institute of Technology Kharagpur 721302, India}
\author{Arup Chakraborty and Indra Dasgupta}
\email{sspid@iacs.res.in}\affiliation{Department of Solid State Physics, Indian Association for the Cultivation of Science, Jadavpur, Kolkata-700032,  India}
\author{Elena Hasanu, Daniele Ercolani and Lucia Sorba}
\email{lucia.sorba@nano.cnr.it}\affiliation{NEST-Istituto Nanoscienze-CNR and Scuola Normale Superiore, Piazza S. Silvestro 12, I-56127 Pisa, Italy}
\author{Mauro Gemmi}
\affiliation{Center for Nanotechnology Innovation @ NEST, Istituto Italiano di Tecnologia, Piazza S. Silvestro 12, I-56127 Pisa, Italy}


\begin{abstract}
We study band alignment in wurtzite-zincblende polytype InAs
heterostructured nanowires using temperature dependent resonance
Raman measurements. Nanowires having two different wurtzite
fractions are investigated. Using visible excitation wavelengths in
resonance Raman measurements, we probe the electronic band alignment
of these semiconductor nanowires near a high symmetry point of the
Brillouin zone (E$_{1}$ gap). The strain in the crystal structure, as
revealed from the shift of the phonon mode,
explains the observed band alignment at the wurtzite-zincblende
interface. Our experimental results are further supported by
electronic structure calculations for such periodic
heterostructured interface.
\end{abstract}

\maketitle
\def\d{{\mathrm{d}}}

\section{Introduction}The electronic structure of III-V As-based semiconductors
in zincblende (ZB) phase are well studied in the literature
\cite{Chelikowsky}. The modulation of the crystal structure in the
wurtzite (WZ) phase of these semiconductors alters various physical
parameters, particularly those related to their optical properties
\cite{Dacal,De}. More intriguing electronic band structure and
related optical characteristics can be achieved by growing
heterostructure of ZB and WZ phases in these systems. In the
literature, the electronic band structure of WZ-ZB heterocrystalline
III-V semiconductor interface has been calculated using
first-principles density functional theory within local-density
approximation \cite{Murayama,Belabbes}. It is believed that the
WZ/ZB heterostructure forms a type-II band alignment with positive
valence band (VB) and conduction band (CB) offset, where the top of
the valence band is defined by the WZ part while the bottom of the
conduction band is hosted by the ZB portion of the heterostructure
\cite{Belabbes}.

In recent years, the polytype crystal structure of III-V
semiconductor  nanowires (NWs) has been exploited in band gap
engineering of these systems for optoelectronic device fabrication
\cite{Joyce}. In polytype NWs, in general, one finds a periodic
quantum well structure of the ZB phase as a narrow segment into the
WZ phase \cite{Caroff}. While governing the optical characteristics
of these systems, their electronic structure may not follow our
understanding on the basis of a single heterostructured interface.
The effect of periodicity is expected to be reflected in the band
alignment, in general, and in the band offset, in particular. Heiss
et. al. have demonstrated that the optical characteristics of the
NWs with periodic arrangement of ZB-WZ phases along the axis are
quite different from those in non-periodic interface {\cite{Heiss}.

There are quite a few report, in which the energy landscape of
polytype heterostructured NWs has been explored using luminescence
(photoluminescence (PL)/ cathodoluminescence (CL)/time-resolved PL)
measurements, thereby discuss the band alignment near the $\Gamma$  point of
the band structure \cite{Heiss,Vainorius,Jahn,Moller1,Spirkoska}.
These studies indicate that the optical transitions in GaAs or in
InAs ZB-WZ heterostructure NWs are mostly governed by (i) band to
band transitions within individual WZ and ZB phases
\cite{Moller1,Spirkoska,Akopian}, (ii) defect related donor-acceptor
recombinations \cite{Vainorius,Jahn,Moller1}, (iii) band-acceptor
transitions \cite{Jahn} and (iv) transition of spatially separated
electrons and holes, i.e. when they are located in the different
sections of ZB/WZ phases of the heterostructure \cite{Jahn,Akopian,Smith}.
It has been demonstrated that other factors, like
the effect of ZB and WZ fractions along the axis of the wires
\cite{Heiss,Spirkoska} and spontaneous polarization over the
valence band \cite{Jahn,Jacopin},  play crucial role in determining
the optical transitions in such systems. The band bending due to
induced space charge by spatially separated electrons and holes,
typical in type II quantum well, has also been discussed in the
literature \cite{Spirkoska,Bao}. In polytype NWs inclusion of
stacking fault/twinning in the crystal structure, the twin density
and periodicity are shown to be reflected in optical transitions in
these systems \cite{Heiss,Spirkoska,Ahtapodov}. To explain the
impact of such disorder in the band alignment  the role of
confinement of either holes or electrons in narrow segments of the
NWs \cite{Heiss,Spirkoska} has been exploited.

Another important aspect, namely the effect of interfacial strain
between ZB and WZ phases on the electronic structure of these NWs
hardly received any attention. There are quite a few reports in the literature, in which the strain in
the crystal of GaAs and InAs polytype NWs has been discussed \cite {Zardo, Panda}. It is well known that the effect of
strain in the crystal structure modifies the electronic band
structure appreciably and cannot be ignored in a related discussion.
Cheiwchanchamnangij and Lambrecht have shown the modulation of the
band structure parameters with lattice strain in WZ and ZB phases of
GaAs using self-consistent GW calculations
\cite{Cheiwchanchamnangij}.

Strain in a crystal structure can be
modulated by temperature. As the phonon spectra are sensitive to
strain in the lattice, we have carried out temperature dependent resonance Raman (RR)
measurements to  demonstrate the effect of strain across the WZ-ZB
interface of InAs polytype NWs in modulating the band alignment. Our experimental studies
are corroborated with electronic structure calculations for periodic
 heterostructures of WZ-ZB interface,  providing further
credence to the important role of strain.
For visible excitation wavelengths RR
spectra of InAs probe the high symmetry A point for WZ phase and L
point of the ZB phase in the bulk electronic band structure. The
corresponding electronic transitions at these points reflect the E$_{1}$
gap of these semiconductors. Here we would like to mention that the optical
band gap (E$_0$ gap) of InAs  in ZB and WZ phases are 0.37 eV \cite{Chelikowsky} and 0.48 eV \cite{De}. Hence,
it is not possible for conventional photoluminescence measurements to probe its electronic band structure.

The remainder of the paper is organized as follows. In section II we
discuss experimental and theoretical methods that we have employed
in the present work. Section III is devoted to the description and
discussion of our experimental and theoretical results followed by
conclusions in section IV.

\section{Methodology}
Aligned InAs NWs were grown on InAs (111)B substrate using chemical
beam epitaxy technique by employing different growth protocols in
order to realize NWs with different WZ/ZB fractions (i.e. different
residual strain) along the NW axis. Two different samples namely NW1
and NW2 were grown at temperatures of 400$\pm$5$^{\circ}$C,
430$\pm$5$^{\circ}$C, respectively
and  metal-organic (MO) line pressure of  trimethyl indium (TMIn) of 0.3
Torr. Instead,  tertiarybutyl arsine (TBAs) was varied to 0.5 Torr and 2.0
Torr for NW1 and NW2 respectively. TEM observations were carried out
on a Zeiss Libra 120 transmission electron microscope operating at
120 kV and equipped with an in-column omega filter for energy
filtered imaging. The NWs were transferred on a formvar carbon
coated copper grid by gently rubbing the grids on the as-grown
substrate.

Stokes micro-Raman measurements were carried out in back scattering
geometry using a spectrometer, equipped with a microscope (model
BX41 Olympus, Japan), triple-monochromator(model T64000, JY, France)
and a Peltier cooled CCD detector. Several excitation lines between
476.486 nm and 568.188 nm from Ar$^{+}$-Kr$^{+}$ laser (Model 2018-RM,
Newport, USA) were used for RR measurements. For room temperature
measurements, the incident laser beam was focused on the sample
using 100$\times$ (numerical aperture 0.9) objective lens and for
temperature dependent measurements 50$\times$ (numerical aperture 0.5)
objective lens was used. Power was kept at 100 kW/cm$^{2}$ to avoid the
effect of laser heating. Room temperature RR spectra, reported in
this article, are corrected for incident laser power, the $\omega^{4}$ law,
optical properties of the material and spectral response of the
experimental set up for different excitation energies \cite{Loudon}.
We assumed the optical parameters of InAs in WZ phase to be the same
as that of the bulk ZB phase. To carry out temperature dependent
Raman measurements, a sample cell (Model THMS600, Linkam Scientific
Instruments, UK) was used. We waited for at least 30 minutes to
stabilize the set temperature before each measurement. To
investigate the variation in band alignment with temperature, we
recorded Raman spectra at different temperatures for seven different
wavelengths between 476.486 nm (2.607 eV) and 568.188 nm (2.186 eV).
We assumed that the optical properties of InAs do not change
significantly over the temperature range of our
interest \cite{Carles1}.
For Raman measurements, the NWs were transferred on a Si substrate.

Electronic structure calculations were carried out in the frame-work
of density functional theory (DFT) as implemented in Vienna
ab-initio Simulation Package (VASP) \cite{Kresse1,Kresse2}. We have
used plane wave basis set with energy cut-off of 500 eV along with
projector augmented wave (PAW) method \cite{Bloch}. Since local
density approximation (LDA) and generalized gradient approximation
(GGA) underestimate the band gap therefore to reproduce the
experimental band gap for the bulk system,  the
Heyd-Scuseria-Ernzerhof (HSE06) \cite{Heyd1,Heyd2}, hybrid functional
for exchange-correlation potential with different fraction of
non-local Fock exchange were used. It is observed that experimental
band gap can be reproduced for the non local Fock exchange  =0.18.
In view of the fact that HSE calculations are computationally
expensive, generalized gradient approximation (GGA) with
Perdew-Burke-Ernzerhof (PBE) functional \cite{Perdew} was used for
the electronic structure calculations of NW heterostructure
with large number of atoms in the unit cell.

\section{Results and Discussion}

\subsection{Polytypism in NWs/Crystal structure of NWs}

\begin{figure}
\centerline{\epsfxsize=3.5in\epsffile{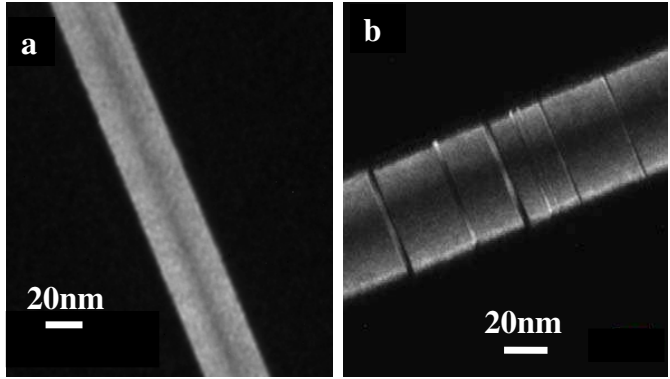}} \caption{Dark
field TEM images of InAs NW segments (a) NW1  (b) NW2 recorded along
the [2-1-10] WZ zone axes.}
\end{figure}

The crystal structure of InAs NWs with different WZ/ZB fractions has
been characterized by dark field transmission electron microscope
(DFTEM) images taken along [2-1-10] WZ (parallel to [110] ZB) zone
axis. In this orientation, ZB and WZ crystal structures show two
different diffraction patterns in which the spots belonging to one
phase are well separated from those of the other. Therefore, a DFTEM
image, obtained by selecting the spots of one of the two phases,
gives a contrast in which stacking faults and polytype changes can
be easily identified. Characteristic DFTEM images of two different
NWs designated as NW1 and NW2 are shown in Fig. 1. The average
fraction of WZ phase, estimated by analyzing DFTEM images of at
least 20 NWs for each sample, are (99$\pm$1)\% and (80$\pm$4)\% for NW1 and
NW2, respectively. The diameters of NW1 and NW2 are 31$\pm$2 nm and 56$\pm$3
nm, respectively. The corresponding lengths are 2.0$\pm$0.2 $\mu$m and
2.5$\pm$0.2$ \mu$m.

\begin{figure}
\centerline{\epsfxsize=2.8in\epsffile{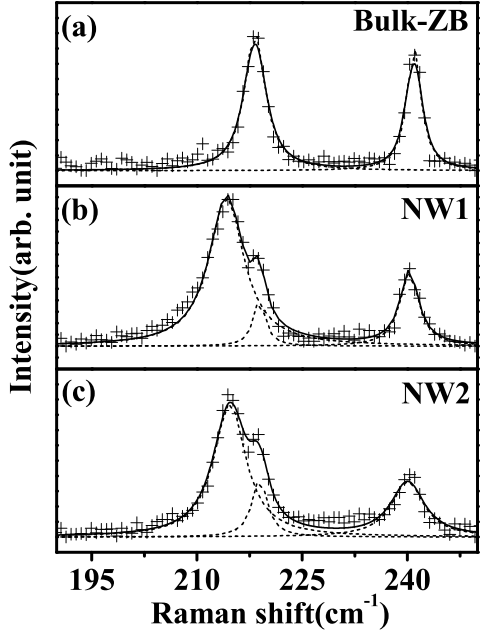}} \caption{Non-resonant Raman
spectra recorded at 100K for (a) bulk, (b) NW1 (99\% WZ), and (c) NW2 (80\% WZ).
Experimental data points are shown by + symbols and the net fitted
spectra by solid lines. The deconvoluted phonon modes are shown by
dashed lines.}
\end{figure}

Fig. 2 shows non-resonant Raman spectra of bulk InAs in ZB phase and
NW1, recorded at 100K using 496.507 nm as the excitation wavelength.
To assign the phonon modes, we deconvoluted each spectrum with
Lorentzian functions, keeping the Raman shift, width and intensity
of all phonon modes as free fitting parameters. The deconvoluted
components are shown by dashed lines in Fig. 2. For bulk InAs in ZB
phase the TO and LO phonon modes appear at 217 cm$^{-1}$ and 239 cm$^{-1}$,
respectively \cite{Carles2}. For NW1 in predominantly WZ phase, an
additional E$_{2}^{H}$ mode at 213 cm$^{-1}$ appears in the spectrum due to
L-point folding of the Brillouin zone of the ZB phase in the
modified crystal structure \cite{Arguello, Zardo}. In Fig. 2(c), the non-resonant
spectrum for NW2  with 80\% WZ fraction is found to be very similar
to that obtained for NW1.

\subsection{Observation of strain in the crystal structure of NWs}

\begin{figure}
\centerline{\epsfxsize=3in\epsffile{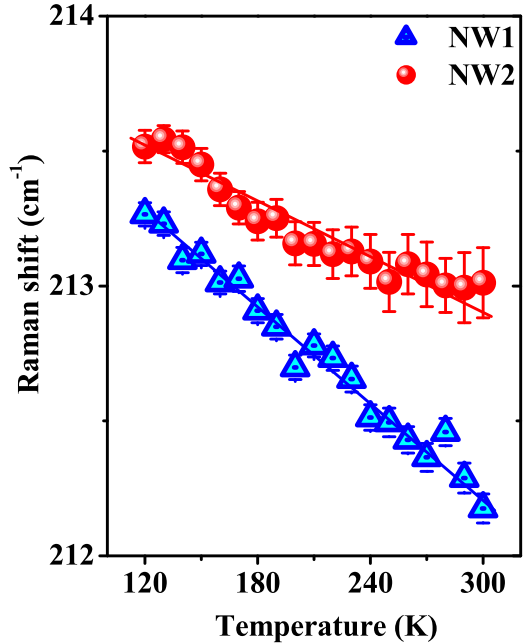}}
\caption{Evolution of Raman shift of the TO phonon
mode with temperature for NW1 and  NW2.}
\end{figure}

The strain in a crystal structure is reflected in phonon wavenumber. Fig. 3 compares the variation of
the Raman shift of the TO phonon mode ($\omega_{TO}$), over the temperature range between 120K and 300K
for NW1 and NW2. Due to the increase in anharmonicity in the crystal
structure, one expects a decrease in $\omega_{TO}$ with temperature
\cite{Balkanski}, as observed in Fig. 3.  However, we find that
the value of
the phonon frequency is appreciably higher in case of NW2 than in
NW1. It is reasonable to assume that the crystal structure of NW1, with
99\% WZ phase, is nearly strain free. Hence, we attribute the
observed higher values of $\omega_{TO}$, obtained for NW2
(in comparison with NW1, see Fig. 3), to be due to
the net compressive strain in its crystal structure. It is to be recalled
(refer to Fig. 1) that in NW2 there are large number of ZB segments
within the WZ segments of the NW. There is an appreciable
difference in lattice parameters between the two phases of InAs
\cite{Zhou} (also refer to the next section) that may aid in
producing strain in NW2. We believe that the
interfacial strain between two phases results in  a higher Raman shift in NW2 than in NW1.  On
the basis of our assumption that the crystal structure of NW1 with
99\% WZ fraction is nearly strain free, the above discussion implies
the existence of strain in the crystal structure of NW2, with 80\%
WZ fraction. Therefore, the effect of strain on the physical
properties related to the crystal structure of these NWs cannot be
ignored.

\subsection{Bulk electronic band structure of InAs}

\begin{figure}
\centerline{\epsfxsize=6in\epsffile{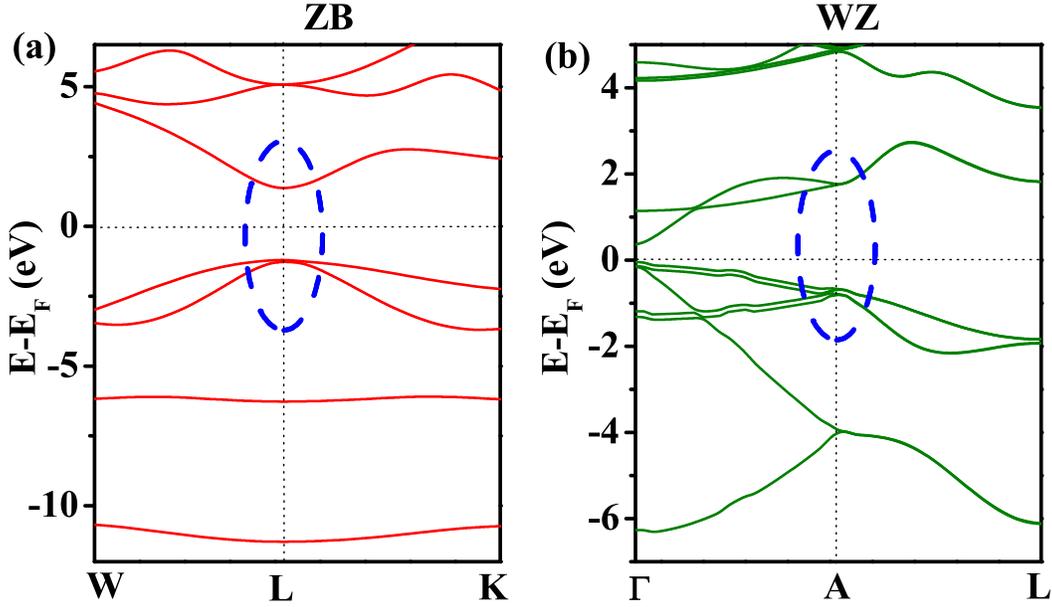}}
\caption{Calculated electronic energy band diagram near E$_{1}$ gap at 0K
of (a) ZB-InAs and (b) WZ-InAs, as obtained from DFT calculations
using hybrid functionals. The high symmetry points, discussed in
this article, are marked by blue dashed lines.}
\end{figure}

Next we calculate electronic band structures of InAs in both
ZB and WZ phases, using hybrid functional method as implemented in
density functional theory (DFT).  Fig. 4 (a) and (b) present the calculated
band structures in these two phases. In our calculation,
we used the lattice parameter a=6.040 \AA\ for the ZB phase
\cite{Tsay}.  We estimate the fundamental band gap for ZB-InAs at 0K to be 0.378 eV.
The E$_{1}$ gap (band gap near the L point
of the Brillouin zone) has been evaluated as 2.568 eV. The lattice
parameters for WZ structure are obtained from that of the ZB
structure using the relation
\begin{equation}
a_{WZ}=\frac{a_{ZB}}{\sqrt{2}}
\end{equation}
and
\begin{equation}
c_{WZ}=\sqrt{\frac{8}{3}}a_{WZ}
\end{equation}
The estimated lattice parameters for the WZ phase are $a$=4.2709 \AA\
and $c$=6.9744 \AA. For the WZ phase, the E$_{1}$ gap (A point of the
Brillouin zone) at 0K is estimated to be 2.432 eV. We find the
difference in calculated E$_{1}$ gap-energies of two phases to be $\sim$136
meV.
\begin{figure}
\centerline{\epsfxsize=2.5in\epsffile{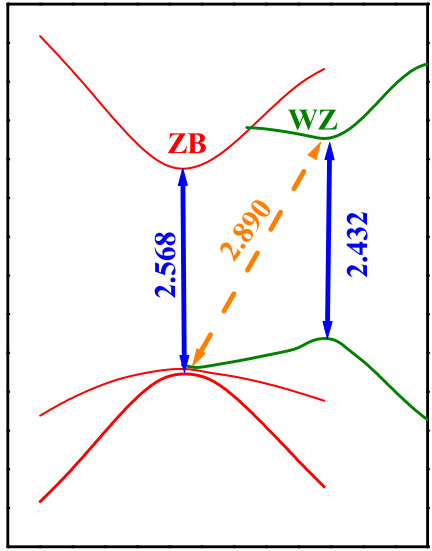}}
\caption{Band alignment
at high symmetry point at WZ-ZB heterointerface (as obtained from
the green marked regions in 4 (a) and (b)). All possible transition
energies (in eV) are also marked. The solid arrows indicate band to
band transitions and the dashed arrow corresponds to transition of spatially separated carriers.}
\end{figure}

In order to obtain the E$_{1}$ band alignment at the ZB-WZ
heterointerface near the high symmetry point (corresponding to L and
A points of the Brillouin zone of ZB and WZ phases respectively), we
consider the portion of the band structure marked by blue dashed lines in Fig.
4(a) and 4(b). The schematic band alignment near WZ-ZB
heterointerface, thus obtained, is shown in Fig. 5 and is of
type-II, where the valence band maximum and conduction band minimum
are located in WZ and ZB region respectively. This alignment of the
E$_{1}$ gap is very similar to that obtained for the E$_{0}$ gap for a ZB-WZ
interface. As mentioned earlier, in the absence of any defect
related states in type II band alignment of a quantum well one
expects either localized band to band electronic transitions in the well
and in the barrier regions (blue arrows in Fig. 5), or transition
forming spatially separated holes and electrons (orange arrow). The
corresponding calculated transition energies from the bulk band
structure are 2.568 eV, 2.432 eV and 2.890 eV, respectively.

\subsection{Electronic structure as obtained from resonance Raman measurements}

Light does not interact directly with phonons in a crystal. The
incident photon excites the electron from its ground state and
creates an electron-hole pair. This electron or hole interacts with
a phonon and creates a new excited state. Next, the electron and
hole recombines and we get the scattered photon.  When the value of
the excitation energy coincides with that of the electronic band
gap, resonance occurs. Thus, RR measurements probe the electronic
band structure of semiconductors. Raman spectroscopy exploits the
coupling of different phonon modes with the electronic states,
following the selection rules. Governed by Raman selection rules, RR
scattering reveals symmetries of the electronic states.

\begin{figure}
\centerline{\epsfxsize=3in\epsffile{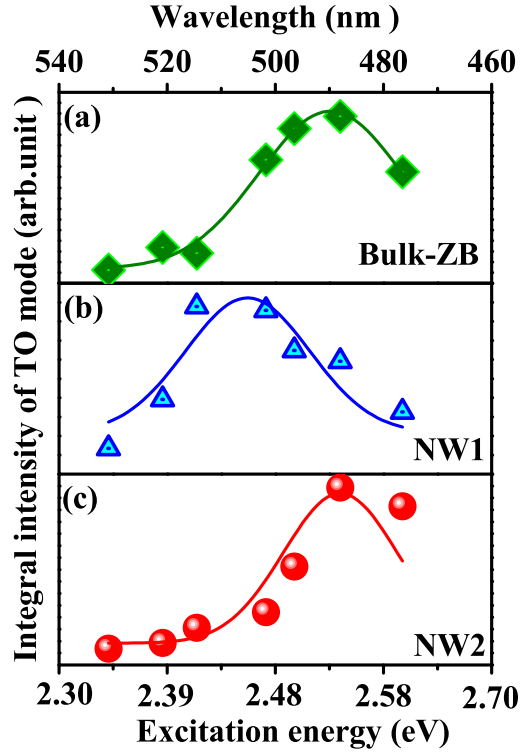}}
\caption{Variation in  intensity of the TO phonon mode at room temperature
for different excitation energies  (a) bulk InAs, (b) NW1 and
(c) NW2.}
\end{figure}
We have carried out RR measurements to probe the band alignment in
NWs. For comparison, we also study RR spectra of bulk InAs, in the
usual ZB phase. For RR measurement, Raman spectra of bulk InAs and
NWs are recorded at different excitation energies (over the range
between 2.340 and 2.607 eV). It is to be noted that the given range
of excitation energies covers a wide range of the visible spectrum
(530 nm to 568 nm) and RR scattering occurs, coupling the energy
states of the E$_{1}$ gap of InAs.  Fig. 6(a)-6(c) present the variation
in intensity of the strong TO phonon mode for bulk and NWs for different
excitation energies. The resonance energies at 2.56$\pm$0.02 eV,
2.46$\pm$0.02 eV and 2.55$\pm$0.02 eV are observed for bulk, NW1 and NW2,
respectively.  For NW1 with 99\% WZ fraction the resonance energy is
~110 meV less than that of bulk InAs of ZB phase,
which is close to the calculated (136 meV) difference in the E$_{1}$ gap
of  ZB and WZ structure of InAs. These values match the earlier report   \cite{Moller2} on difference in E$_1$ gap of InAs in ZB and WZ phases fairly well. Interestingly, for NW2 with 80\% WZ
fraction, the resonance energy is observed to be 2.55 eV, a value
close to that of InAs in the ZB phase.

Keeping in mind that interfacial strain can modify the band
alignment and to explain the observed difference in resonance energy
in NW1 and NW2, we propose a possible explanation for the observed
resonance energy of NWs and hence, plausible band alignment at the
heterointerface.  For nearly strain-free NW1 with 99\% WZ fraction,
Raman resonance occurs at 2.46 eV due to minimum energy band to band
transitions at the WZ segment of the wire (see Fig. 5). The band
gap in the ZB segment remains at 2.56 eV (as observed in bulk
ZB-InAs). As the ZB fraction in this set of wires is only 1\%, the
phonon creation is dominated by band to band transition of electrons
in the WZ segment. The spatially separated transition (ZB$\rightarrow$WZ) could not
be observed, since the corresponding energy is beyond the range of
excitation energies used by us.

\begin{figure}
\centerline{\epsfxsize=2.5in\epsffile{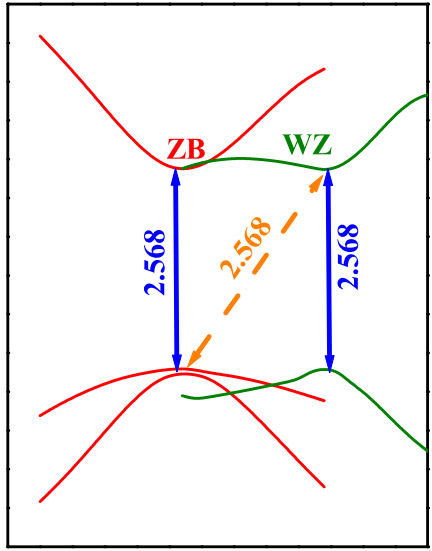}}
\caption{Expected band alignment at heterointerface of NW2.}
\end{figure}

To anticipate the possible band alignment in NW2, we refer to Fig.
3, where we have plotted the variation of $\omega_{TO}$  with
temperature. We attributed the observed differences in the phonon
characteristic of NW1 and NW2 to be a consequence of compressive strain
in the crystal structure of the latter. For resonance energy of NW2
to be the same as that of InAs of ZB phase, one of the possibilities
is a lowering of both valence and  conduction band edge of the WZ phase, so that
they match with those of the ZB
phase. The proposed band alignment in NW2 is shown in Fig. 7.
It is to be noted that
for strain-induced shift in the conduction or valence band edge of
the ZB segment the resonance
would have occurred at different excitation energies, not exactly at
2.55 eV, which corresponds to a gap of the ZB phase. Further, if
 the conduction (valence) band edge of the WZ phase lies below (above) the one
we show schematically in Fig. 7, then the energy gap of the WZ
phase will be less than the lowest probing excitation energy, used
by us. In that case, the observed resonance is still expected to be
governed by the band to band transition of the ZB phase.

\subsection{Theoretical justification for strain induced band alignment}

\begin{figure}[h]
\centerline{\epsfxsize=4.5in\epsffile{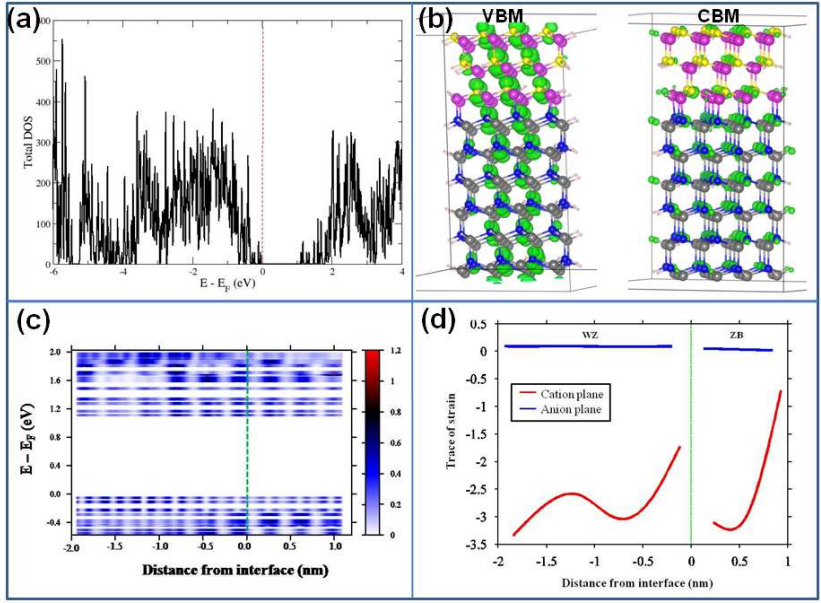}}
\caption{(a)Density of states (DOS) for the periodic ZB-WZ coupled NW with
30\% ZB part; (b) Iso-surface (in green color) of charge density for
VBM and CBM of the periodic coupled NWs (30 \% ZB);
In(WZ), As(WZ), In(ZB), As(ZB), H atoms are in grey, blue, magenta,
yellow, and, sky blue color respectively. (c) energy resolved charge
density plot for the  periodic coupled NWs
with 30\% ZB part. (d) strain profile across the interface for the
coupled ZB-WZ periodic NWs.}
\end{figure}

In order to understand the band alignment due to ZB-WZ interface in
NW2, we constructed a model of coupled WZ-ZB NW from bulk supercell
structure that closely resemble the crystal structure of NW2. In the
model structure, the WZ NW consists of 72 In and 72 As atoms
and ZB NW consists of 31 In and 31 As atoms.  As a consequence
the diameter of both the systems are 1.4 nm while their lengths are
1.82 nm and 0.77 nm respectively. The dangling bonds at the surface
of the NWs are saturated by pseudo hydrogen atoms
\cite{Huang}. The concentration of the ZB part in this model
NW heterostructure is 30\%. The (111) As terminated polar
facet of ZB is attached to the (0001) polar facet of WZ containing
In atoms. We have considered the system to be periodic (-ZB-WZ-ZB-)
along its length. In order to reduce the cost of computation, although
the calculations are done for 30\% ZB, our results are expected also
to hold good for 20\% ZB NWs realized experimentally.

The density of states (DOS) for the periodic
coupled NW (see Fig. 8(a)) reveal that the gap between valence
band maximum (VBM) and the conduction band minimum (CBM) to be 1.15
eV. Interestingly the band gap for a  WZ NW (diameter = 1.4
nm) is 1.18 eV while that for  ZB NW of same diameter is
calculated to be 1.14 eV, suggesting that the band gap for the
coupled ZB-WZ periodic NW is very close to that obtained for the ZB
system. As mentioned earlier (Section IIIc), this band alignment for the E$_{0}$ gap is also
expected to hold good for the E$_{1}$ gap. Our
calculations of the charge density (see Fig. 8(b)) and energy
resolved charge density (see Fig. 8(c)) further reveal that for
the periodic system there is no band offset at the interface. The
plot of the charge density for the valence band maximum (VBM) and the
conduction band minimum (CBM)(see Fig. 8(b)) show that the charge density
is delocalized and spread both in the ZB and WZ part of the coupled NW. The
energy resolved charge density plot shown in Fig. 8(c) clearly
reveal the absence of band offset both in the valence band and the
conduction band. Finally we have calculated the strain profile for
the coupled quantum NW following the method proposed in ref. \cite{Ganguli}
where the trace of the strain tensor that represents the volumetric
strain is calculated for the system. The results of our calculation
for the cationic and anionic planes are displayed in Fig. 8(d). The strain in the anion
 plane is tensile while the strain in the cation plane is compressive. In
contrast to coupled quantum dots \cite{Ganguli}, we not only find appreciable
strain at the interface but also substantial strain even far away
from the interface. The strain profile is nearly constant for the anion plane but
oscillating for the cation plane. Such large value of strain is expected to have
profound impact on the band alignment as anticipated from the
experimental results discussed in the previous section.

\subsection{Temperature variation of the band alignment}
\begin{figure}[h!]
\centerline{\epsfxsize=6in\epsffile{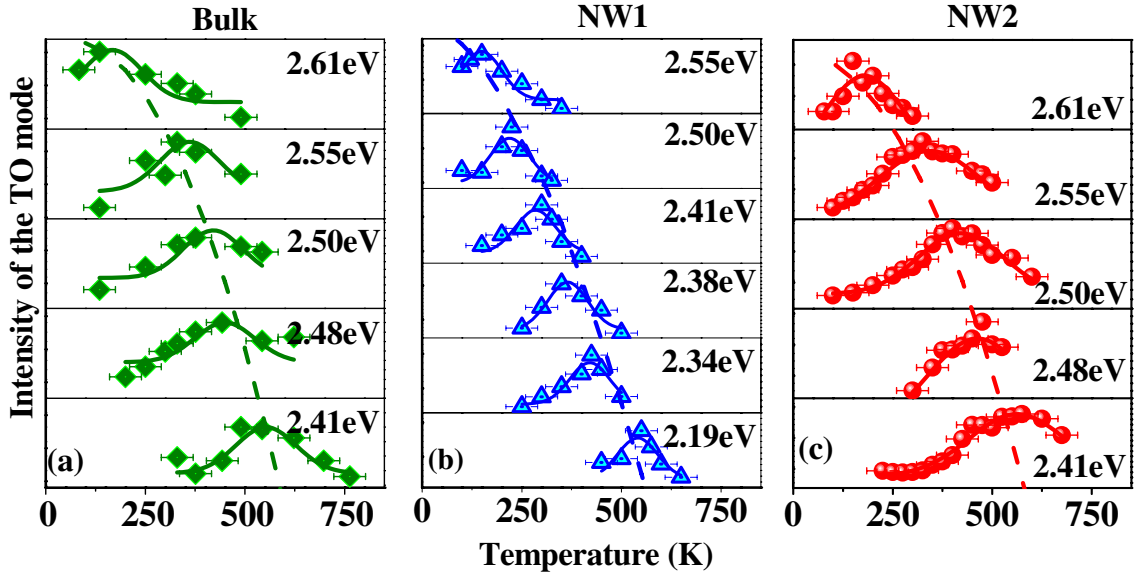}}
\caption{Intensity of the TO phonon mode at different temperatures for
different excitation energies   (a) bulk InAs, (b) NW1 and (c) NW2. The
solid lines are guide to the eyes. }
\end{figure}
The strain in a crystal is sensitive to the temperature. We have
probed the change in the band gaps of NW1 and NW2 with temperature.
We recorded Raman spectra of both bulk and NWs at different
temperatures for various excitation energies. The spectra are
recorded at different temperature range between 120K and 750K for
each excitation wavelength. The intensity of the TO phonon mode
for each spectrum was estimated. The data points in Fig. 9 (a), (b) and (c)
correspond to the integral intensity of the TO phonon mode at the
given temperature for bulk-InAs, NW1 and NW2, respectively.  The dashed lines indicate the
shift in temperature at which resonances were obtained at different
excitation energies. In Fig. 10 we present the variation of the
resonance energy with temperature for bulk InAs and NWs, as obtained
from Fig. 9 (plots the maximum of the temperature-response curve in
each panel in Fig. 9 (a), (b) and (c)).  It is to be recalled that
by RR measurements we probe the E$_{1}$ gap of InAs and thus, the above
measurements reveal the temperature variation of the high symmetry
point in the band structures of the bulk and the NW systems.

\begin{figure}
\centerline{\epsfxsize=3in\epsffile{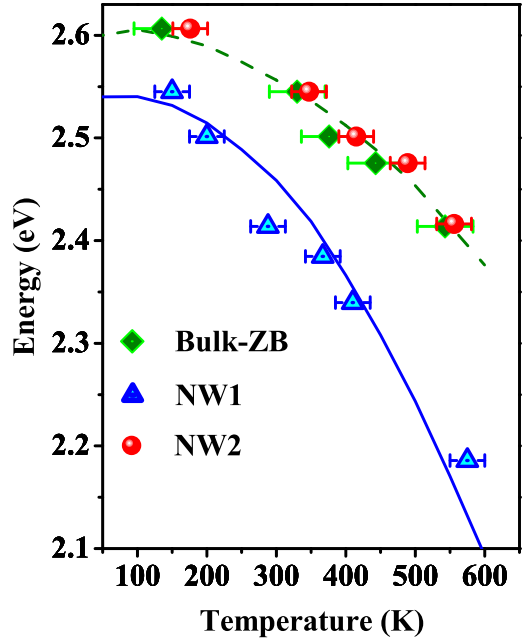}}
\caption{Variation in resonance energy with temperature of bulk, NW1 and  NW2. The
calculated (scaled) variation of the E$_{1}$ gap with temperature for WZ and ZB phases are shown by blue solid line and green dashed line, respectively.}
\end{figure}

We have calculated the variation of the E$_{1}$ gap for the WZ and the ZB
with temperature. For the ZB phase, the values of the lattice
parameter at different temperatures are obtained from Ref. \cite{Tsay}.
Due to unavailability of the lattice parameters of WZ-InAs at
different temperatures, we calculated the temperature dependent
lattice parameters for the WZ phase using a$_{ZB}$ and the relations
described in Eqn.1 and Eqn.2. The variation in E$_{1}$ gap of ZB and WZ phases
with temperature, as obtained from the DFT+hybrid calculations, are
shown in Fig. 10 by green (dashed) and blue (solid) lines, respectively. The
theoretically estimated E$_{1}$ vs T plot are scaled to match with the
experimental data points.  We find that though the variation of the
resonance energy with temperature in NW1 is like one expects for the
WZ phase, the same for NW2 is similar to that of the ZB phase. Thus,
we independently confirm the dominating ZB-like behavior of the band
alignment in NW2 for a wide range of temperature from 120K to 750K
under visible light.

\section{Conclusion}

In this article we correlate the band alignment in InAs WZ-ZB
heterostructured NWs with different WZ fractions, with the lattice
strain in their crystal structure. While RR measurements probed the
high symmetry points of the electronic band structure of InAs, the
strain in the crystal structure is revealed from the shift of the phonon modes.  Theoretical
simulation on model system confirm the dominating ZB like behavior
for the band alignment for NW2 with 20\% ZB fraction. We have
demonstrated the ZB-like behavior of the band alignment in InAs
WZ-ZB polytype NWs with 20\% ZB fraction, over a wide temperature
range.
\section{Acknowledgement}
ID thanks DST India and Cray Super Computer facility in IACS Kolkata.



\begin{thebibliography}{66}

\bibitem{Chelikowsky}J. R. Chelikowsky and M. L. Cohen, Phys. Rev. B \textbf{14}, 556 (1976).

\bibitem{Dacal} L. C. O. Dacal and A Cantarero, Material Research Express \textbf{1},
015702 (2014).

\bibitem{De} A. De and C. E. Pryor, Phys. Rev. B \textbf{81}, 155210 (2010).

\bibitem{Murayama} M. Murayama and  T. Nakayama, Phys. Rev. B \textbf{49}, 4710 (1994).

\bibitem{Belabbes}A. Belabbes, C. Panse, J.  Furthmï\"{u}ller, and F.  Bechstedt, Phys.
Rev. B \textbf{86}, 075208 (2012).


\bibitem{Joyce}H. J. Joyce, Q. Gao, H. H. Tan, C. Jagadish, Y. Kim, J. Zou, L. M.
Smith, H. E. Jackson, J. M.Yarrison-Rice, P. Parkinson and M. B.
Johnston, Progress in Quantum Electronics \textbf{35}, 23 (2011).

\bibitem{Caroff}P. Caroff, K. A. Dick, J. Johansson, M. E. Messing, K. Deppert, and
L. Samuelson, Nat. Nanotechnol. \textbf{4}, 50 (2009).

\bibitem{Heiss}M. Heiss, S. Conesa-Boj, J. Ren, H.-H. Tseng, A. Gali, A. Rudolph,
E. Uccelli, F. Peir\'{o}, J. R. Morante, D. Schuh, E. Reiger, E.
Kaxiras, J. Arbiol and A. Fontcuberta i Morral, Phys. Rev. B \textbf{83},
045303 (2011).

\bibitem{Vainorius} N. Vainorius, D. Jacobsson, S. Lehmann, A. Gustafsson, K. A. Dick,
L. Samuelson, and M-E. Pistol, Phys. Rev. B \textbf{89}, 165423 (2014).

\bibitem{Jahn} U. Jahn, J. L\"{a}hnemann, C. Pf\"{u}ller, O. Brandt, S. Breuer, B.
Jenichen, M. Ramsteiner, L. Geelhaar, and H. Riechert, Phys. Rev. B
\textbf{85}, 045323 (2012).

\bibitem{Moller1} M. Mï\"{o}ller, M. M. de Lima Jr, A. Cantarero, T. Chiaramonte, M. A.
Cotta and F. Iikawa, Nanotechnology \textbf{23}, 375704 (2012).

\bibitem{Spirkoska}D. Spirkoska, J. Arbiol, A. Gustafsson, S. Conesa-Boj,  F. Glas,
I. Zardo, M. Heigoldt,  M. H. Gass, A. L. Bleloch, S. Estrade, M.
Kaniber, J. Rossler, F. Peiro, J. R. Morante, G. Abstreiter, L.
Samuelson, and A. Fontcuberta i Morral, Phys. Rev. B \textbf{80}, 245325
(2009).

\bibitem{Akopian}N. Akopian, G. Patriarche, L. Liu, J.-C. Harmand, and V. Zwiller,
Nano. Lett. 10, 1198 (2010). 12L. M. Smith, H. E. Jackson, J. M.
Yarrison-Rice, and C. Jagadish, Semicond. Sci. Technol. \textbf{25}, 024010
(2010).

\bibitem{Smith}L. M. Smith, H. E. Jackson, J. M. Yarrison-Rice, and C. Jagadish,
Semicond. Sci. Technol. \textbf{25}, 024010 (2010).



\bibitem{Jacopin}G. Jacopin, L. Rigutti, L. Largeau, F. Fortuna, F. Furtmayr, F. H. Julien, M. Eickhoff,
and M. Tchernycheva, J. Appl. Phys. \textbf{110}, 064313 (2011).


\bibitem{Bao} J. Bao, D. C. Bell, F. Capasso, J. B. Wagner, T. Martensson, J. Tr\"{a}grdh,
 and L. Samuelson, Nano. Lett. \textbf{8}, 836  (2008).



\bibitem{Ahtapodov} L. Ahtapodov, J. Todorovic, P. Olk, T. Mjaland, P. Sl\.{o}ttnes, D. L. Dheeraj,
A. T. J. van Helvoort, B-O. Fimland, and H. Weman, Nano. Lett. \textbf{12},
6090 (2012).

\bibitem{Zardo}I. Zardo, S. Conesa-Boj, F. Peiro, J. R. Morante, J. Arbiol, E.
Uccelli, G. Abstreiter, and A. Fontcuberta i Morral, Phys. Rev. B
\textbf{80}, 245324 (2009).

\bibitem{Panda}J. K. Panda, A.  Roy, A. Singha, M. Gemmi, D. Ercolani, V.
Pellegrini and L. Sorba, Appl. Phys. Lett. \textbf{100}, 143101 (2012).

\bibitem{Cheiwchanchamnangij}T. Cheiwchanchamnangij and W. R. L. Lambrecht, Phys. Rev. B \textbf{84},
035203 (2011).


\bibitem{Loudon}R. Loudon, J. Phys. \textbf{26}, 677 (1965).

\bibitem{Carles1}R. Carles, N. Saint-Cricq, J. B. Renucci, A. Zwick, and M. A.
Renucci, Phys. Rev. B  \textbf{22}, 6120 (1980).

\bibitem{Kresse1}G. Kresse and J. Hafner, Phys. Rev. B \textbf{47}, 558 (1993).

\bibitem{Kresse2}G. Kresse snf J. Furthmuller, Phys. Rev. B \textbf{54}, 11169 (1996).

\bibitem{Bloch}P. E. Bloch, Phys. Rev. B \textbf{50}, 17953 (1994).

\bibitem{Heyd1}J. Heyd, G. E. Scuseria, and M, Ernzerhof, J. Chem. Phys. \textbf{118}, 8207
(2003).

\bibitem{Heyd2}J. Heyd, G. E. Scuseria, and M, Ernzerhof, J. Chem. Phys. \textbf{124},
219906 (2006).

\bibitem{Perdew}J. P. Perdew, K. Burke, and M. Ernzerhof, Phys. Rev. Lett. \textbf{77}, 3865
(1996).

\bibitem{Carles2}R. Carles, N. Saint-Cricq, J. B. Renucci, M. A. Renucci, and A. Zwick, \textbf{22}, 4804 (1980).




\bibitem{Arguello}C. A. Arguello, D. L. Rousseau, and S. P. S. Porto, Phys. Rev. \textbf{181},
1351 (1969).

\bibitem{Balkanski}M. Balkanski, R. F. Wallis and E. Haro, Phys. Rev. B \textbf{28}, 1928
(1983).


\bibitem{Willardson}R. K. Willardson, Semiconductors and Semimetals (Vol 2), Physics of
III-V Semiconductors (Academic, New York, 1966), p. 171.


\bibitem{Zhou}F. Zhou, A.L. Moore, J. Bolinsson, A. Persson, L. Fr�berg, M.T.
Pettes, H. Kong, L. Rabenberg, P. Caroff, D. A. Stewart, N. Mingo,
K. A. Dick, L. Samuelson, H. Linke, and L. Shi, Phys. Rev. B  \textbf{83},
205416 (2011).



\bibitem{Tsay}Y. F. Tsay, B. Gong, S. S. Mitra, and J. F. Vetelino, Phys. Rev. B
\textbf{6}, 2330 (1972).

\bibitem{Moller2}M. M\"{o}ller, M. M. de Lima Jr., and A. Cantarero, Phys. Rev. B \textbf{84}, 085318 (2011).

\bibitem{Huang}X. Huang, E. Lindgren, and J. R. Chelikowsky, Phys. Rev. B 71, 165328 (2005).



\bibitem{Ganguli}N. Ganguli, S. Acharya, and I. Dasgupta, Phys. Rev. B \textbf{89}, 245423
(2014).


\end{thebibliography}
\end{document}